\documentclass[5p,twocolumn]{elsarticle}
\usepackage{graphicx,latexsym}
\usepackage{dcolumn}
\usepackage{amssymb,amsmath,bm}
\usepackage{subfigure}
\usepackage{braket}
\usepackage{siunitx}
\usepackage{array}
\usepackage{float}
\usepackage[nopar]{lipsum}
\usepackage{caption}
\usepackage[super]{nth}

\newcommand{\angstrom}{\text{\normalfont\AA}}
\usepackage{hyperref}
\hypersetup{
	pdfnewwindow=true,       
	colorlinks=true,         
	linkcolor=blue,          
	citecolor=blue,          
	filecolor=magenta,       
	urlcolor=black           
}

\usepackage[normalem]{ulem}

\def\sec#1{Sec.\ \ref{#1}}

\def\fig#1{Fig.\ \ref{#1}}

\journal{}

\begin{document}

\begin{frontmatter}


\title{Electronic, thermal, and optical properties of graphene like SiC$_x$ structures: \\
	  Significant effects of Si atom configurations}

\author[a1,a2]{Nzar Rauf Abdullah}
\ead{}
\address[a1]{Division of Computational Nanoscience, Physics Department, College of Science, 
	University of Sulaimani, Sulaimani 46001, Kurdistan Region, Iraq}
\address[a2]{Computer Engineering Department, College of Engineering, Komar University of Science and Technology, Sulaimani 46001, Kurdistan Region, Iraq}

\author[a1]{Gullan Ahmed Mohammed}

\author[a1]{Hunar Omar Rashid}    
  
\author[a6]{Vidar Gudmundsson}   
\address[a6]{Science Institute, University of Iceland,
	Dunhaga 3, IS-107 Reykjavik, Iceland}

%

\begin{abstract}
We investigate the electronic, thermal, and optical characteristics of graphene like SiC$_x$ structure using model calculations based on density functional theory. The change in the energy bandgap can be tuned by the Si atomic configuration, rather than the dopants ratio. The effects of the concentration of the Si atoms and the shape of supercell are kept constant, and only the interaction effects of two Si atoms are studied by
varying their positions. If the Si atoms are at the same sublattice positions, 
a maximum bandgap is obtained leading to an increased Seebeck coefficient and figure of merit. 
A deviation in the Wiedemann-Franz ratio is also found, and a maximum value of the Lorenz number is thus discovered.
Furthermore, a significant red shift of the first peak of the imaginary part of the dielectric function towards the visible range of the electromagnetic radiation is observed. On the other hand, if the Si atoms are located at different sublattice positions, a small bandgap is seen because the symmetry of sublattice remains almost unchanged. Consequently, the Seebeck coefficient and the dielectric function are only slightly changed compared to pristine graphene. In addition, the electron energy loss function is suppressed in Si-doped graphene. These unique variations of the thermal and the optical properties of Si-doped graphene are of importance to understand experiments relevant to optoelectronic applications.
\end{abstract}


\end{frontmatter}

\section{Introduction}

The discovery of carbon-related materials and their development has been considered to be crucial for the future of technology \cite{Akinwande2019,doi:10.1021/nl102824h, doi:10.1063/1.3676276, Lee385, Shahil2012, Zhang2005}.   
Carbon based material such as monolayer graphene offers good possibility of device performance down to the atomic limit \cite{doi:10.1002/admi.201700232}.
Extraordinary properties of graphene make it an interesting material for many electronic devices. The zero band gap of graphene leads it to be a metallic like material with nontrivial physical properties such as high electron mobility at room temperature and exceptional thermal conductivity \cite{YU201576}. The bandgap of graphene can be tuned by several techniques termed functionalization strategies \cite{Sinha2019}, doping \cite{Choi2010}, application of an electric field \cite{PhysRevB.80.195401} or a magnetic field \cite{doi:10.1063/1.3457353}, or strain \cite{MONTEVERDE2015266}, etc.
Among these methods, the doping process is considered one of the most
feasible ways to tune the bandgap of graphene and altering its physical
characteristics, for the reason that doping can break or alter the symmetric structure of
graphene \cite{C0JM02922J}.

The integration of two-dimensional graphene into silicon chips promises a better heterogeneous platform to deliver a massively enhanced potential based on silicon technology \cite{Kim2011}.
Silicon doped graphene systems (Si$_x$C$_y$) have emerged as new 2D materials called siligraphenes. They have been considered as good composite materials based on graphene 
to study enormously interesting physical application such as solar cell devices \cite{doi:10.1021/nl403010s} and gas sensors \cite{Dong2017}. 
Si-substituted graphenes have been identified as significant structures 
for technological application \cite{doi:10.1002/adma.201403537}. The ratio of the Si doping in graphene determines the potential or the functionality of 
the structure. For instance, SiC$_2$ can be used to built solar cell devices due to its finite band gap \cite{doi:10.1021/nl403010s}, 
SiC$_3$ and SiC$_5$ can be used as topological insulators and semi-metals \cite{Dong2017, PhysRevB.89.195427}, while SiC$_7$ is suggested for 
photovoltaic devices \cite{Dong2016,Abdullah_2018}.

In recent years, the investigation of thermal properties of graphene like SiC$_x$ structures (g-SiC$_x$) has been attracting the attention of many researchers, and the interest in the subject is still growing \cite{Lee2015,C8RA06156D}. 
It has been shown that the electrical and thermal properties of two configurations
of siligraphene, g-SiC$_3$ and g-SiC$_7$ can be tuned by the concentration of 
the silicon. The thermal conductivity of g-SiC$_7$ is exponentially enhanced with
temperature, but it varies parabolically for g-SiC$_3$ \cite{Houmad2018, Houmad2019}.
Composite thermal interface materials have been optimized with an admixture
of graphene structures with different doped materials leading to a high enhancement 
of the effective thermal conductivity \cite{Shahil2012}, which has been beneficial for large range of materials used by industry.
Therefore, graphene based materials can be used for energy harvesting and increased 
efficiency of solar cell devices at room temperature due to a high-recorded thermal 
property dominated by the acoustic phonons \cite{Balandin2011}.

Optical properties of g-SiC$_x$ is another important aspect of research in which the g-SiC$_x$ has been used for solar cell application. Optical properties such
as the optical conductivity, reflectivity and refractive index have been studied and it has been demonstrated that the optical conductivity is enhanced due to 
the Si-dopant \cite{Houmad2015,ABDULLAH20181432}. 
Theoretical studies have shown that the bandgap of g-SiC$_x$ is decreased
with increasing applied external electrical field intensity. 
The dielectric functions and the refractive indexes at low frequency are 
decreased compared with those of pristine graphene \cite{C9RA00326F}.
Furthermore, the optical absorption spectra of a Si-doped graphene are 
dominated by bound Frenkel exciton and the Si-dopant enhances the optical conductivity 
of graphene in the visible range \cite{SHAHROKHI20171185}.

In this work, we model Si-doped graphene nanosheets with a $6.25\%$ concentration 
ratio of Si-atoms.
We will show the influence of Si atomic configurations, and the Si-Si interactions on 
the electronic, thermal and optical characteristics using model calculations based on Density Functional Theory (DFT) \cite{RASHID2019102625,ABDULLAH2020126350,Nzar_arXive_2020}.

In \sec{Sec:Model} the structure of Si-doped graphene nanosheets is briefly over-viewed. In \sec{Sec:Results} the main achieved results are presented and analyzed. In \sec{Sec:Conclusion} the conclusions are presented and summarized.

\section{Computational Tools}\label{Sec:Model}

We consider two-dimensional Si-doped sheet of graphene with $32$ atoms consisting of $30$ carbon and $2$ silicon atoms. It is configured into a $4\times4\times1$ supercell with $6.25\%$ Si doping concentration ratio. The electronic properties of the systems are calculated within a self-consistent field (SCF) approximation. We use a Kohn-Sham local density approximation \cite{PhysRev.140.A1133} augmented a with generalized gradient approximation (GGA),
as implemented by the Quantum Espresso (QE) package for the band structure and the density of states of the systems \cite{Giannozzi_2009}. 

The Brillouin zone is sampled using a $\Gamma$-centered $12\times12\times1$ $k$-point grid 
and atomic relaxations are continued until the Helmann-Feynman forces acting
on the atoms is less than \SI{0.02}{eV\per{\angstrom}} and the total energy changes are less than $10^{-6}$~eV. In addition, the convergence criteria of the SCF and the non-SCF of density of state calculations are set to be $10^{-3}$~eV with an $8\times8\times1$, and a $100\times100\times1$ $k$-point grid, respectively. 
A plane-wave projector-augmented wave (PAW) method with a kinetic energy cut-off 
equal to $680.28$~eV is implemented in the DFT code \cite{PhysRevLett.77.3865, PhysRevB.59.1758, PhysRevB.50.17953, ABDULLAH2018223}.
Furthermore, the exchange-correlation function calculated by the non-relativistic Perdew Burke Ernzerhof (PBE) approach is utilized in our model using the QE package \cite{Petersen2000}.

The calculations of the thermal properties of the systems are carried out using 
the Boltzmann theory implemented in the BoltzTraP package \cite{Madsen2006,abdullah2019manifestation}. The BoltzTraP code 
uses a mesh of band energies and is well documented within the QE guidelines. We calculate the Seebeck coefficient, $S$, the figure of merit, $ZT$, of the systems. 
In the BoltzTraP, the Seebeck coefficient is given by \cite{PhysRevB.77.165119}

\begin{equation}
S = \frac{ek_{\mathrm{B}}}{N\Omega} \sigma^{-1} \int d\varepsilon \Big(-\frac{\partial f_0}{\partial \varepsilon}\Big) \Big(\frac{\varepsilon - \mu}{k_{\mathrm{B}} T}\Big) \, \gamma_{n,\kappa},
\end{equation}
where 
\begin{equation}
\gamma_{n,\kappa} = \sum_{n,\kappa} \tau_{n,\kappa} \,  \vec{v}_{n,\kappa} \vec{v}_{n,\kappa} \, \delta(\varepsilon-\varepsilon_{n,\kappa}).
\end{equation}
Herein, $e$ is the unit charge, $k_{\mathrm{B}}$ is the Boltzmann's constant,
$N$ indicates the number of $k$-points, $\Omega$ refers to the unit cell volume,
$\sigma$ is the electrical conductivity, $\varepsilon$ is the band energy, $f_0$ represents Fermi-Dirac distribution function,  $\mu$ stands for the chemical potential, $T$ is the temperature measured in Kelvin, $\tau$ is the relaxation time, $v$ is the group velocity of the charges, and $\delta$ is the Dirac delta function. The subscripts $\kappa$ and $n$ mean the crystal momentum and the band index~\cite{Madsen2006,Abdullah2019}.

In addition, the $ZT$ is defined by the following equation:
\begin{equation}
ZT = \dfrac {S^{2} \, \sigma}{k_{th}} \, T,
\end{equation}
where $k_{th} = k_{\rm e} + k _{\rm p}$ is the sum of both the electron and the phonon
thermal conductivity. 

The imaginary part of dielectric function $\varepsilon_{\text{imag}}(\omega)$, in the long wavelength limit, can be
obtained directly from the electronic structure calculation
\begin{align}
\varepsilon_{\text{imag}}(\omega) & =  \Big(\frac{4\, \pi e^2}{\omega^2 m^2}\Big)\sum_{i,j} \int \braket{i| M_{cv}|j}^2 \, f_i(1-f_j) \nonumber \\
& \times \delta(E_f - E_i - \omega) \, d^3k ,
\end{align}
where $M_{cv}(k)$ are the dipole transition matrix elements \cite{KHENATA200629}.

\section{Results}\label{Sec:Results} 

\begin{figure*}[htb]
	\centering
	\includegraphics[width=1.0\textwidth]{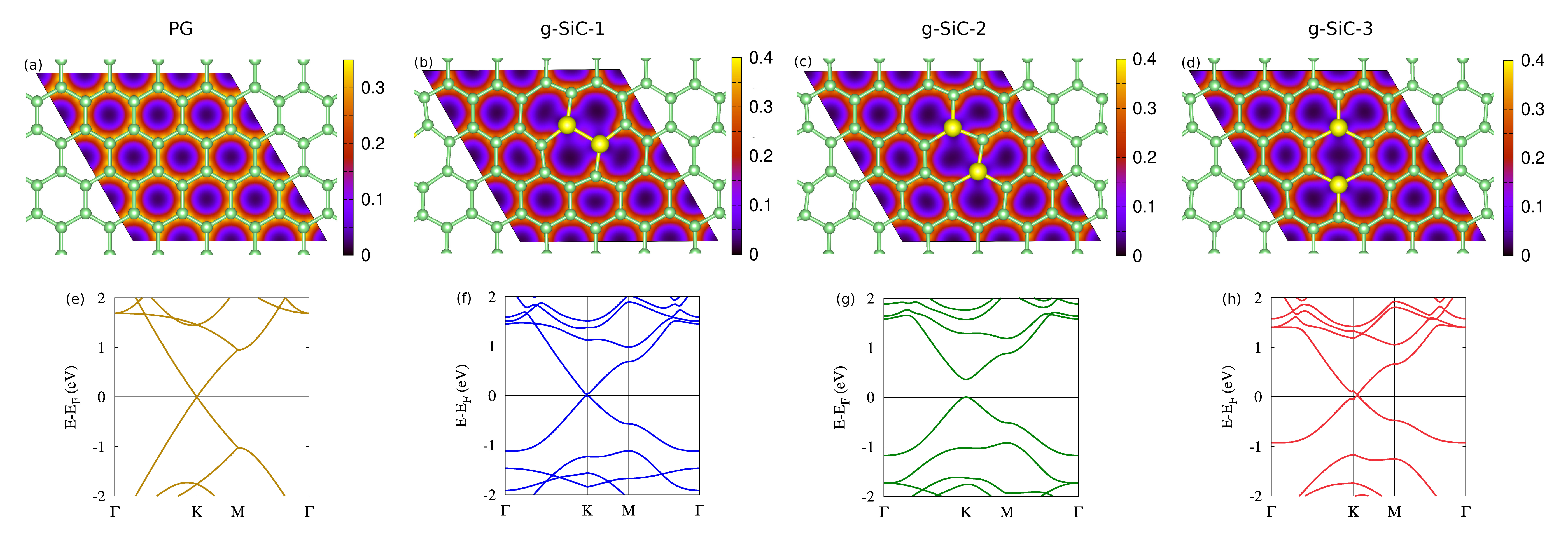}
	\caption{The $4\times4\times1$ supercell of PG (a), g-SiC-1 (b),  g-SiC-2 (c), and  g-SiC-3 (d) structures with their electron charge distribution (contour plot) are plotted. The corresponding band structure of four structures are shown from (e) to (h).}
	\label{fig01}
\end{figure*}

In this section, we present the main results calculated via DFT. 
We consider three Si doped graphene structures in addition to pure graphene (PG) 
as shown in \fig{fig01}(a-d). The three Si-doped graphene structures are: 
First, the two Si atoms (yellow) are doped at a para- and an ortho-position which means that the Si atoms are adjacent. The structure is identified as a g-SiC-1 (b). So, the two Si atoms are at different sublattice positions in which one dopant atom is placed at sublattice A and the other is at B. 
Second, the two Si atoms are put at the para- and the meta-position, $i.\,e.$  
the Si atoms are at the same sublattice positions (either A or B), forming g-SiC-2 (c).
\lipsum[0]
Finally, the two Si atoms are both located at a para-position, resulting in g-SiC-3, shown in (d), in this case the two Si atoms are at different lattice positions~\cite{C2RA22664B, WEI20141841}.

Among these Si-doped structures, g-SiC-1 seems to be less structurally stable 
compared to g-SiC-2 and g-SiC-3 since sp$^2$-hybridization is not favored by Si–Si bonds,
a planar structure should be energetically unfavorable~\cite{Li2014}.

The electron charge distributions of the four structures are plotted in \fig{fig01} (contour plots). The high electron charge distribution between the carbon atoms in the PG 
indicates strong and stiff covalent bonds forming between the C atoms. 
Therefore, PG boasts of great stability and a very high tensile strength (the force with which one can stretch something before it breaks) \cite{Kashanieaat6951}.
In the g-SiC-1 structure, the Si atoms form Si-Si bonds and the electron charge 
distribution between the Si atoms is very weak because of the presence a strong repulsive 
force between the Si atoms.
Furthermore, the C-Si bonds are somewhat polarized towards the C atoms due to carbon's greater electronegativity. The electron charge distribution thus concentrates around the C atoms. 

The same scenarios for electron charge distribution are also true for the other two 
Si-doped structures, g-SiC-2 and g-SiC-3. The electron charge distributions of all 
three Si-doped structures confirm that the C-Si bonds are longer and weaker than the
C–C bonds. This could be expected because of the larger 
atomic radius of a Si compared to a C atom.

Electronic band structure is calculated along the high-symmetry 
$\Gamma$-K-M-$\Gamma$ directions.
By looking at the band structures, we can see that the conduction and valence bands 
touch at the Fermi level, zero energy, as is presented in \fig{fig01}(e) for PG. 
One can see from the band structures in \fig{fig01}(f–h) that the
maximum band gap appears for g-SiC-2 and it is equal to $0.358$~meV.
The category of g-SiC-2 originates when both the Si dopants are 
placed at the same sublattice positions (either A or B). 
This observation confirms that the origin of the bandgap is due to symmetry breaking of
graphene sublattices, which is maximized in these configurations. 
The same scenario has been seen for Boron or Nitrogen doped graphene \cite{C2RA22664B}.

\subsection{Thermal properties}

We display thermoelectric properties at $100$~K which is in the intermediate temperature 
range, $50\text{-}160$~K, where the electron and the lattice temperatures are decoupled. 
The electronic thermal conductivity $k_e$ is thus proportional to the charge conductivity 
at a given temperature. 
In this temperature range, the electron and the lattice temperatures are very well
decoupled in low-disorder graphene \cite{doi:10.1021/nl403967z, PhysRevB.83.205421}.

\begin{figure}[htb]
	\centering
	\includegraphics[width=0.35\textwidth]{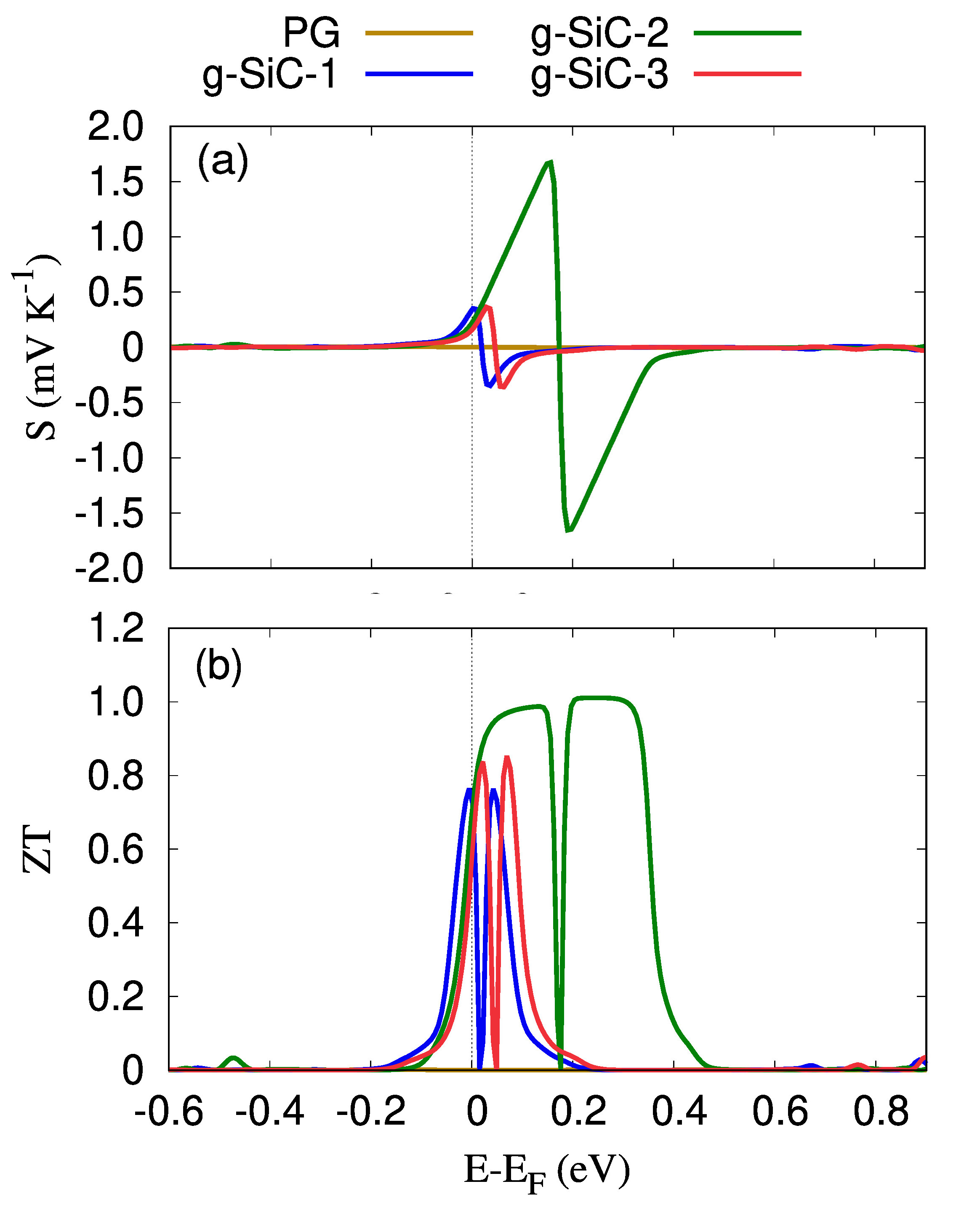}
	\caption{The Seebeck coefficient, $S$, (a) and the figure of merit, $ZT$, (b) at intermediate temperature $T = 100$~K for PG (brown), g-SiC-1 (blue), g-SiC-2 (green) and g-SiC-3 (red).}
	\label{fig02}
\end{figure}

Figure \ref{fig02} shows the contributions of the electrons to different thermoelectric parameters, the Seebeck coefficient (a) and the figure of merit (b).
In general, semiconductor materials with high a bandgap effectively minimize 
free charge carrier contributions. The Seebeck coefficient is thus enhanced since $S$ is inversely proportional to charge carrier concentration \cite{ZOU200182}.
The Seebeck coefficient is found to be significant and as high as $1.67$~mV K$^{-1}$ 
for g-SiC-2 (green line),
$i.\,e.$ higher than the value of 6~$\mu$V K$^{-1}$ for PG.
The enhancement of the Seebeck coefficient is attributed to the
existence of the high bandgap of g-SiC-2 compared to PG. 
In addition, the Seebeck coefficient of g-SiC-1 and g-SiC-3 is smaller than that of g-SiC-2 
due to the smaller bandgaps of these two structures.

Figure of merit is what ultimately would determine the efficiency of the devices. 
It is interesting to compare the $ZT$ of the four aforementioned structures.
One can see that increased Seebeck coefficient raises the enhancement of figure of merit as is presented in \fig{fig02}(b), where the maximum $ZT$ is found for g-SiC-2 (green line). 

\begin{figure}[htb]
	\centering
	\includegraphics[width=0.35\textwidth]{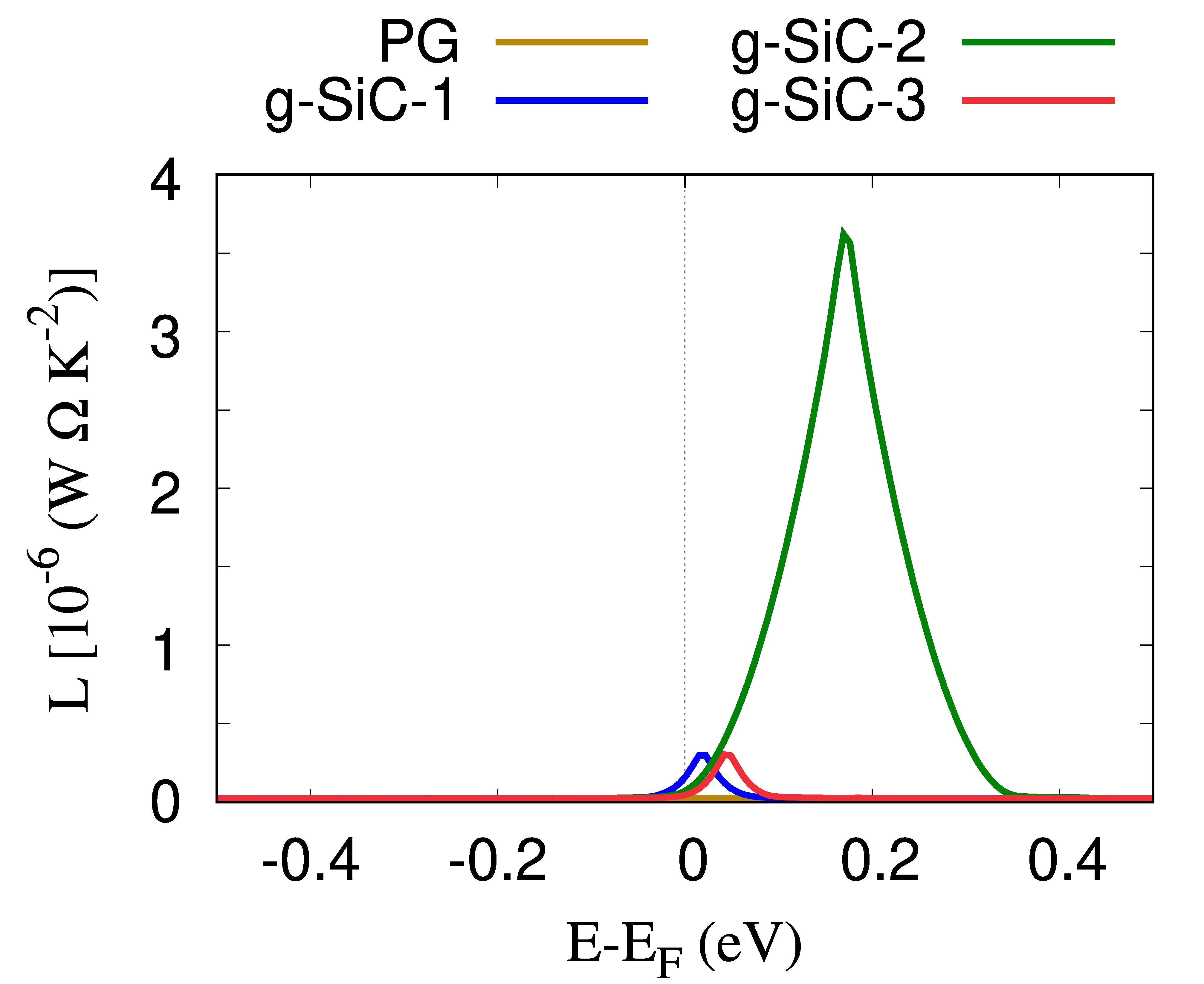}
	\caption{The Lorenz number, $L$, at the intermediate temperature $T = 100$~K for PG (brown), g-SiC-1 (blue), g-SiC-2 (green) and g-SiC-3 (red).}
	\label{fig03}
\end{figure}

We now consider the Wiedemann-Franz (WF) law for our model. The WF law  states that the 
ratio of $k_e$ to $\sigma$ is given by
\begin{equation}
\frac{k_e}{\sigma} = L T
\end{equation}
which is constant for ordinary metals, where $L = L_0 = \pi^2 \, k_{\mathrm B}^2/3e^2 = 2.44 \times 10^{-8}$~W $\Omega$ K$^{-2}$, called the Lorenz number \cite{PhysRevB.76.115409, PhysRevB.67.144509}. For graphene, the low chemical potential makes the Lorenz number 
sensitive to resonance scattering and the energy dependence of the relaxation time \cite{PhysRevB.91.115410}. 
Figure \ref{fig03} displays the Lorenz factor for all four structures shown in 
\fig{fig01} at the intermediate temperature $T = 100$~K. 
It can clearly be seen that $L$ strongly depends on the energy or chemical potential, 
especially around the resonance states. In addition,
the value of $L$ depends on the location of the resonance states of the different 
Si atomic configuration in our calculations.
The maximum value of $L=3.62 \times 10^{-6}$~W $\Omega$ K$^{-2}$ is found for g-SiC-2.
We conclude that the WF law is not obeyed in our Si-doped graphene structures at 
$T = 100$~K \cite{doi:10.1021/nl403967z}.

\subsection{Optical properties}

Graphene has unique optical characteristics and can absorb a wide range of electromagnetic radiation \cite{Fan2018}. PG is known to emit light when it has been excited 
by a near-infrared laser \cite{doi:10.1021/nl501376a}. In the case of doped graphene, photoluminescence is observed by creation of a bandgap \cite{doi:10.1063/1.3098358}.

\begin{figure}[htb]
	\centering
	\includegraphics[width=0.35\textwidth]{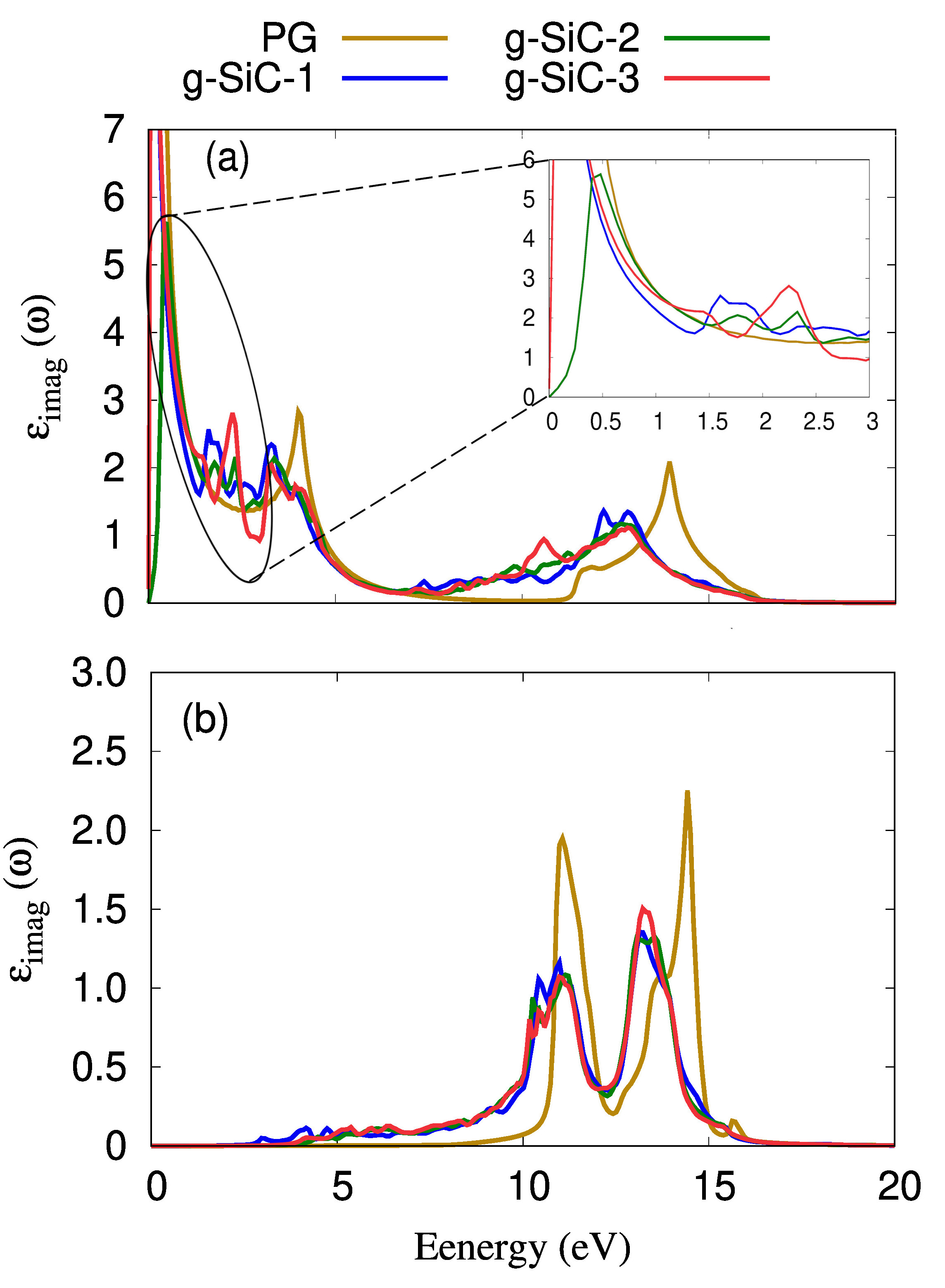}
	\caption{The imaginary part of dielectric function, $\varepsilon_{\text{imag}}(\omega)$, is plotted for PG (brown), g-SiC-1 (blue), g-SiC-2 (green), and g-SiC-3 (red) in the presence of a parallel (a) and a perpendicular (b) electric field.}
	\label{fig04}
\end{figure}

We use the three aforementioned silicon atom configurations to investigate the optical properties of doped graphene. The imaginary part of the dielectric function, 
$\varepsilon_{\text{imag}}(\omega)$, is presented in \fig{fig04} for PG (brown), g-SiC-1 (blue), g-SiC-2 (green), and g-SiC-3 (red) in the case of parallel (a) and perpendicular (b) electric field.
In the case of PG, two peaks in the imaginary part of dielectric function for a 
parallel electric field are observed (brown line), one 
at $4.0$~eV indicating $\pi \rightarrow \pi^*$ transition and another at $14.0$~eV 
displaying $\sigma \rightarrow \sigma^*$ transition. The position of both peaks in our calculation is 
in a good agreement with the literature \cite{PhysRevB.81.155413, PhysRevB.77.233406}.
The intensity ratio of these two peaks is 
$\epsilon_{\text{imag},L}/\epsilon_{\text{imag},R} = 2.82/2.092 = 1.34$ demonstrating that the $\pi \rightarrow \pi^*$ transition is stronger than the $\sigma \rightarrow \sigma^*$ transition. This is what should be obtained for monolayer pure graphene.
In the case of a perpendicular electric filed \fig{fig04}(b), two main peaks for PG appear 
at $11.06$ and $14.42$~eV belonging to the $\pi \rightarrow \sigma^*$ and the $\sigma \rightarrow \pi^*$ transitions, respectively. 

Let's look at the effect of silicon doping of graphene on the imaginary part of dielectric function. 
It is interesting to see an extra peak at $0.4$~eV, in the visible range, in the case 
of a parallel electric field for g-SiC-2 (see the inset of \fig{fig04}(a)) \cite{RANI201428}. The extra peak can be referred to the existence of a finite bandgap.  
Our analysis shows a significant red shift of the first peak of the dielectric function 
towards the visible range. The peak structure becomes more complex due to the 
Si doping, and more peaks may be identified in the visible range.
Furthermore, the intensity of both peaks is decreased in the case of Si-doping for perpendicular electric field as is shown in \fig{fig04}(b). 
This indicates that both transitions, $\pi \rightarrow \sigma^*$ and 
$\sigma \rightarrow \pi^*$, are weaker compared to the corresponding transitions 
active in PG.
These unique variations of the dielectric function of Si doped graphene are of importance to understand relevant experiments and design for optoelectronic applications \cite{WANG20191,doi:10.1002/andp.201700334}.

\begin{figure}[htb]
	\centering
	\includegraphics[width=0.35\textwidth]{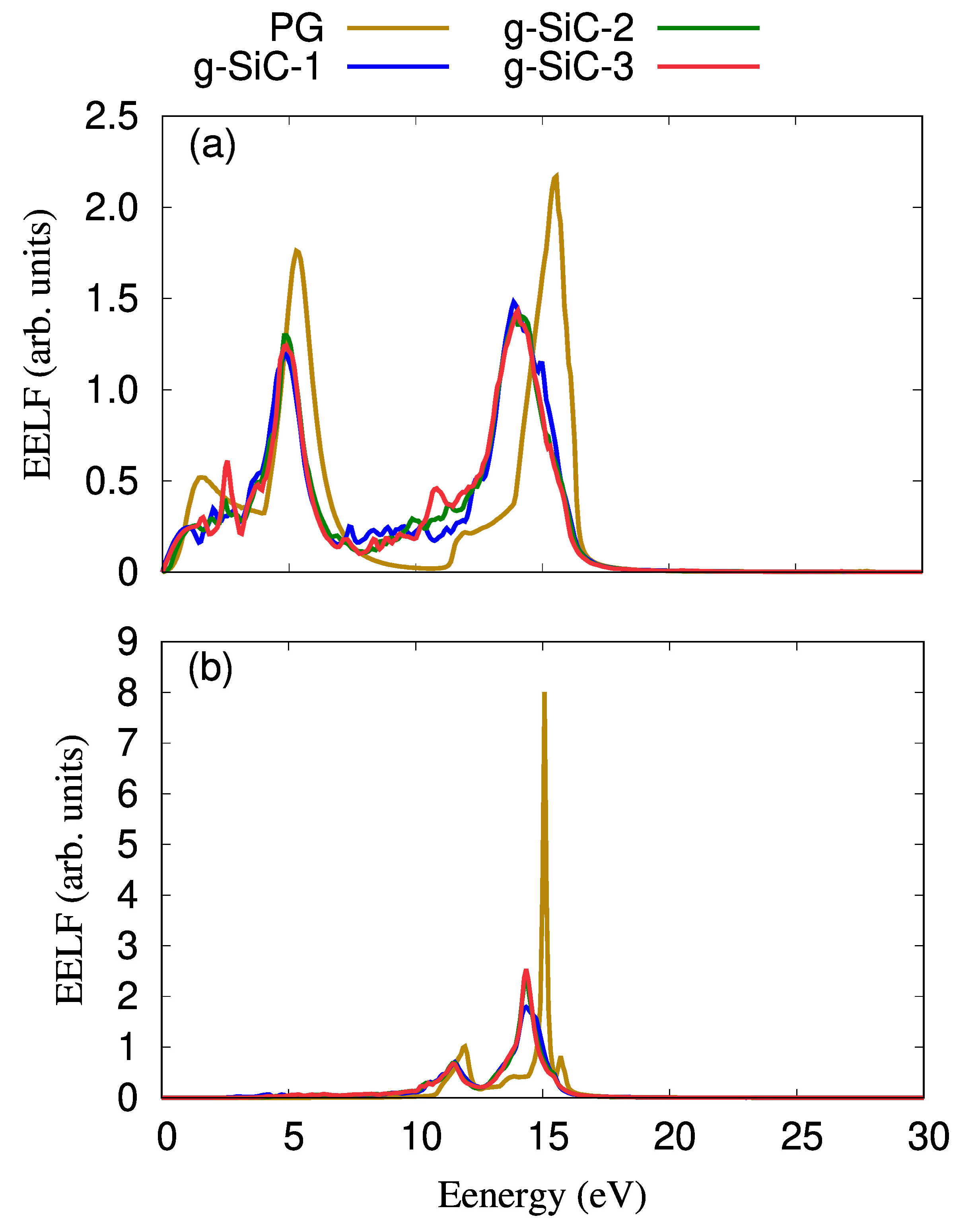}
	\caption{EELF is plotted for PG (brown), g-SiC-1 (blue), g-SiC-2 (green), and g-SiC-3 (red) in the presence of a parallel (a) and a perpendicular (b) electric field.}
	\label{fig05}
\end{figure}

The electron energy-loss function (EELF) is shown in \fig{fig05} in the presence of 
parallel (a) and perpendicular (b) electric field for PG (brown), g-SiC-1 (blue), g-SiC-2 (green), and g-SiC-3 (red). The electron inelastic interaction with a structure is closely related to the EELF which represents the probability of inelastic scattering events.
It can clearly be seen that the EELF is decreased for all three Si-doped graphene structures. 
This is attributed to the low probability of charge concentration close to a Si-dopant in the graphene nanosheets \cite{PhysRevB.77.233406}.   
It should be mentioned that EELS obtains a maximum value where
$\varepsilon_{\text{imag}}(\omega)$ has a finite small value. 
So, plasmonic excitations can be responsible for the maximum intensity of EELS peak \cite{DHAR20171589}.

\section{Conclusion}\label{Sec:Conclusion}

In summary, the characteristics of monolayer graphene-like materials with SiC$_x$ 
stoichiometry are investigated where the position of the Si atoms plays an important role 
for their physical properties. 
The electronic band structure of graphene can be remarkably modulated by different Si doping
configurations, leading to Dirac point shifting and even opening of bandgaps. Consequently 
the figure of merit is enhanced. In addition, the optical properties, such as the dielectric
and the electron energy loss function are influenced by the tuning of the bandgap. 
As a result, a reduction in 
electron energy loss function has been found, and an extra peak in imaginary part of 
the dielectric function towards the visible range is found. 
Last but not least, we show that our model does not obey the  
Wiedemann-Franz law. Therefore, a maximum value of Lorenz number is found around 
resonant scattering states.

\section{Acknowledgment}
This work was financially supported by the University of Sulaimani and 
the Research center of Komar University of Science and Technology. 
The computations were performed on resources provided by the Division of Computational Nanoscience at the University of Sulaimani.


\end{document}